\documentclass[fleqn,usenatbib]{mnras}

\usepackage{newtxtext,newtxmath}
\usepackage[T1]{fontenc}
\usepackage[dvipsnames]{xcolor}
\usepackage{graphicx,adjustbox}
\usepackage{xcolor}
\usepackage[graphicx]{realboxes}
\usepackage{pifont}
\usepackage{mathtools}
\usepackage{comment}
\usepackage{orcidlink}

\title[Constraining the asymmetry of transients from unresolved radio observations]{Constraining inhomogeneities and asymmetries in  SNe, FBOTs, and other high-energy transients from unresolved radio observations}

\author[De Colle et al.]{Fabio De Colle\,\orcidlink{0000-0002-3137-4633}$^{1}$\thanks{E-mail: fabio@nucleares.unam.mx},
Rosa L. Becerra\,\orcidlink{0000-0002-0216-3415}$^{2}$,
Lizeth A. Meza\,\orcidlink{0009-0006-2938-5962}$^{1}$,
Nayana A.J.\,\orcidlink{0000-0002-8070-5400}$^{3,4}$, 
James K.\ Leung\,\orcidlink{0000-0002-9415-3766}$^{5,6,7}$,\newauthor
Luca Izzo\,\orcidlink{0000-0001-9695-8472}$^{8}$,
Raffaella Margutti\,\orcidlink{0000-0003-4768-7586}$^{3,4}$, 
Gerardo Urrutia\,\orcidlink{0000-0002-7834-3113}$^{9}$,
Enrique Moreno-Méndez\,\orcidlink{0000-0002-5411-9352}$^{10}$,\newauthor and Leonardo García-García\orcidlink{0000-0001-5125-1043}$^{11}$\,
\\
$^{1}$Instituto de Ciencias Nucleares, Universidad Nacional Aut{\'o}noma de M{\'e}xico, A. P. 70-543 04510 D. F. Mexico\\
$^{2}$Instituto de Astronom\'ia, Universidad Nacional Aut{\'o}noma de M{\'e}xico, A. P. 70-543 04510 D. F. Mexico\\
$^{3}$Department of Astronomy, University of California, Berkeley, CA 94720-3411, USA\\
$^{4}$Berkeley Center for Multi-messenger Research on Astrophysical Transients and Outreach (Multi-RAPTOR), University of California, Berkeley, CA 94720-3411, USA\\
$^{5}$David A. Dunlap Department of Astronomy and Astrophysics, University of Toronto, 50 St. George Street, Toronto, ON M5S 3H4, Canada\\
$^{6}$Dunlap Institute for Astronomy and Astrophysics, University of Toronto, 50 St. George Street, Toronto, ON M5S 3H4, Canada\\
$^{7}$Racah Institute of Physics, The Hebrew University of Jerusalem, Jerusalem 91904, Israel\\
$^{8}$INAF, Osservatorio Astronomico di Capodimonte, Salita Moiariello 16, I-80131 Napoli, Italy\\
$^{9}$Department of Astronomy and Astrophysics, University of California, Santa Cruz, CA 95064, USA \\
$^{10}$Facultad de Ciencias, Universidad Nacional Aut{\'o}noma de M{\'e}xico, A. P. 70-542, 04510 D. F. Mexico\\
$^{11}$ Instituto de Astronom{\'\i}a, Universidad Nacional Aut\'onoma de M\'exico, km 107 Carretera Tijuana-Ensenada, 22860 Ensenada, Baja California, México
}

\date{Accepted XXX. Received YYY; in original form ZZZ}

\pubyear{2021}

\begin{document}
\label{firstpage}
\pagerange{\pageref{firstpage}--\pageref{lastpage}}
\maketitle

\begin{abstract}
Synchrotron emission, commonly observed in high-energy transients, is produced by relativistic electrons accelerated by shocks. 
As high-energy transients are often unresolved even on angular scales probed by very long baseline interferometry, it is difficult to obtain a full picture of the ejecta and circumstellar medium (CSM) properties that are probed by the radio synchrotron emission. 
Radio spectra of high-energy transients frequently show optically thick slopes shallower than the standard $F_\nu \propto \nu^{5/2}$ expected from synchrotron self-absorption (SSA) models, or broader spectra near the self-absorption frequency. Such deviations are often interpreted phenomenologically, without providing clear insights into the structure of the emitting region.
In this paper, we show how information on the homogeneity and symmetry of the emitting region can be directly inferred from SSA spectra, even when the source is unresolved. We discuss the circumstances under which inhomogeneities in the emitting region can change the low-frequency spectrum (below the self-absorption frequency), even causing it to follow a different slope. We further examine which parameters can be constrained from observations and which remain degenerate. 
We apply this method to the stripped-envelope supernova (SN) 2016coi and to the fast blue optical transient (FBOT) AT2018cow, showing that SSA spectra effectively constrain the degree of inhomogeneity in these systems. We argue that the deviation of the SSA spectrum from the homogeneous expectation provides strong evidence for  inhomogeneities in the emitting region in the SN 2016coi, and asymmetry in the case of AT2018cow, and we infer the characteristics of the emitting region. 
When well sampled spectra are available,  our method can be applied as a general, model-independent, inference method. This approach can be used to constrain inhomogeneities in a variety of unresolved high-energy astrophysical transients, including SNe, FBOTs, tidal disruption events and gamma-ray bursts.
\end{abstract}


\maketitle


\section{Introduction}
\label{sec:introduccion}

High-energy transients, such as gamma-ray bursts, fast blue optical transients (FBOT), supernovae (SNe), and tidal disruption events, among others, produce bright non-thermal emission that traces shocks and provides valuable information on the particle acceleration process, the amplitude, and structure of the magnetic field in the post-shock region, and the density stratification of the environment \citep[see, e.g.,][and references therein]{Gehrels2013,Guidorzi2021}.

The structure of these sources can be probed by very-long-baseline interferometry (VLBI) observations. 
Nevertheless, these sources are typically unresolved due to their large distances from Earth, making it difficult to determine their geometry and structure. 
Even if a source is resolved, it is often only possible to measure the source size, or separate the source into a few different components.

Determining the source geometry and structure is typically done by using techniques such as: i) spectropolarimetry, which measures the polarisation of light and partially reveals the geometry and orientation of the explosion 
\citep[e.g.,][]{Wang2008, Yang2025, Christy2026}, ii) line profile variations,
which probe asymmetries in the ejecta via distortions in line profiles produced by anisotropic velocities in the ejecta, \citep[e.g.,][]{Bevan2016}, iii) radio and X-ray observations, which trace differences in the morphology and intensity of emission due to asymmetry in the explosion 
\citep[e.g.,][]{Chandra2012, Brethauer22, nayana2025b}, and iv) nebular-line-emission and kinematic studies, which map variations in optical and infrared emission lines \citep[e.g.,][]{mazzali2003,
Maeda2010, milisav2015a, milisav2015b}. 

In non-relativistic or weakly relativistic sources  propagating in low-density media, where free-free absorption is negligible, the radio spectrum is typically well described by a synchrotron self-absorption (SSA) model \citep{chevalier98}. In such a model, the low-frequency part of the spectrum rises as $\nu^{5/2}$, and the high-frequency portion falls as $\nu^{-(p-1)/2}$,
$p$ being the power-law index of the accelerated electron population.
In some sources, the low-frequency or the high-frequency slopes are flatter than expected, or the transition between the low- and high-frequency parts of the spectrum is broader than predicted
\citep[e.g.][]{chandra2024b,
soderberg2005, chandra2019, nayana2020, sfaradi2024, nayana2025b}. 
Extreme cases are FBOT AT2018cow \citep{Ho2019, Margutti2019} and SN 2012au (Wiston et al, in preparation), which show clear evolution of the low-frequency spectrum, becoming increasingly flatter over time. 

\citet{bjornsson2013} and \citet{bjornsson2017} showed that the aforementioned spectral features can be explained if the emitting region is inhomogeneous. In their model, they consider a power-law distribution of magnetic field intensities in the emitting region
and introduce a covering factor, which specifies the fraction of the emission area corresponding to a certain  magnetic field intensity. According to the authors, this model provides ``a qualitative discussion of the characteristics of the inhomogeneities that could be deduced from the observations'' \citep{bjornsson2024}.

In this paper, we generalize previous models and provide a detailed discussion of what can be inferred on the inhomogeneities of the source from observations of SSA spectra. 
We begin with a phenomenological approach, describing inhomogeneities in terms of observable quantities, independent of any specific model for the synchrotron radiation. 
We show that the amount of information that can be extracted from unresolved SSA spectra depends on the quality of the data and prior knowledge of the source. When densely sampled spectra are available, a fully general approach can be used to infer the structure of the emitting region.
For a spherical emitting region, i.e., where flux and self-absorption frequencies  depend only on the shock radius $R$, and using a power-law distribution of the magnetic field in each patch, our results reproduce those presented by \citet{bjornsson2013} and \citet{bjornsson2017}.

This paper is organised as follows: In Section~\ref{sec:methods}, we describe the synchrotron self-absorbed emission produced by homogeneous and inhomogeneous sources. Section~\ref{sec:structure} presents a detailed description of the method employed to infer the structure of the emitting region. In Section~\ref{sec:applications}, we show how radio emission can be used to determine the degree of inhomogeneity in SNe and FBOTs. In Section~\ref{sec:discussion} we discuss the results obtained in the paper. 
Finally, in Section~\ref{sec:summarize}, we summarise our findings.

\section{The spectrum of inhomogeneous shocks}
\label{sec:methods}

We consider a population of non-thermal electrons accelerated by a non-relativistic shock. The electrons acquire an energy distribution $N(E) \propto E^{-p}$ for energies $E\geq E_{\rm min}$, where $E_{\rm min}$ 
is the minimum energy of the accelerated electrons, and $p$ is the power-law slope of the electron distribution. Electrons advected through the magnetised post-shock region emit synchrotron radiation. We consider inhomogeneous or asymmetric shocks produced by local fluctuations such as clumps, turbulence, or angular variations in the emitting regions. The  ``standard'' case of a homogeneous emitting region can be obtained as a limiting case. 

We show that there are two approaches to constrain the source asymmetry. In the first approach, when sufficiently dense data  are available, we directly reconstruct the distribution of emitting patches by solving the inverse problem (Section~\ref{model:general}). Alternatively, when only a few data points are available and/or there are strong indications of asymmetry in the emitting regions, we can employ a physically motivated model to gain insights into the properties of the emitting region (Section~\ref{model:mot}).

\begin{figure}
\centering
\includegraphics[width=0.9\linewidth]{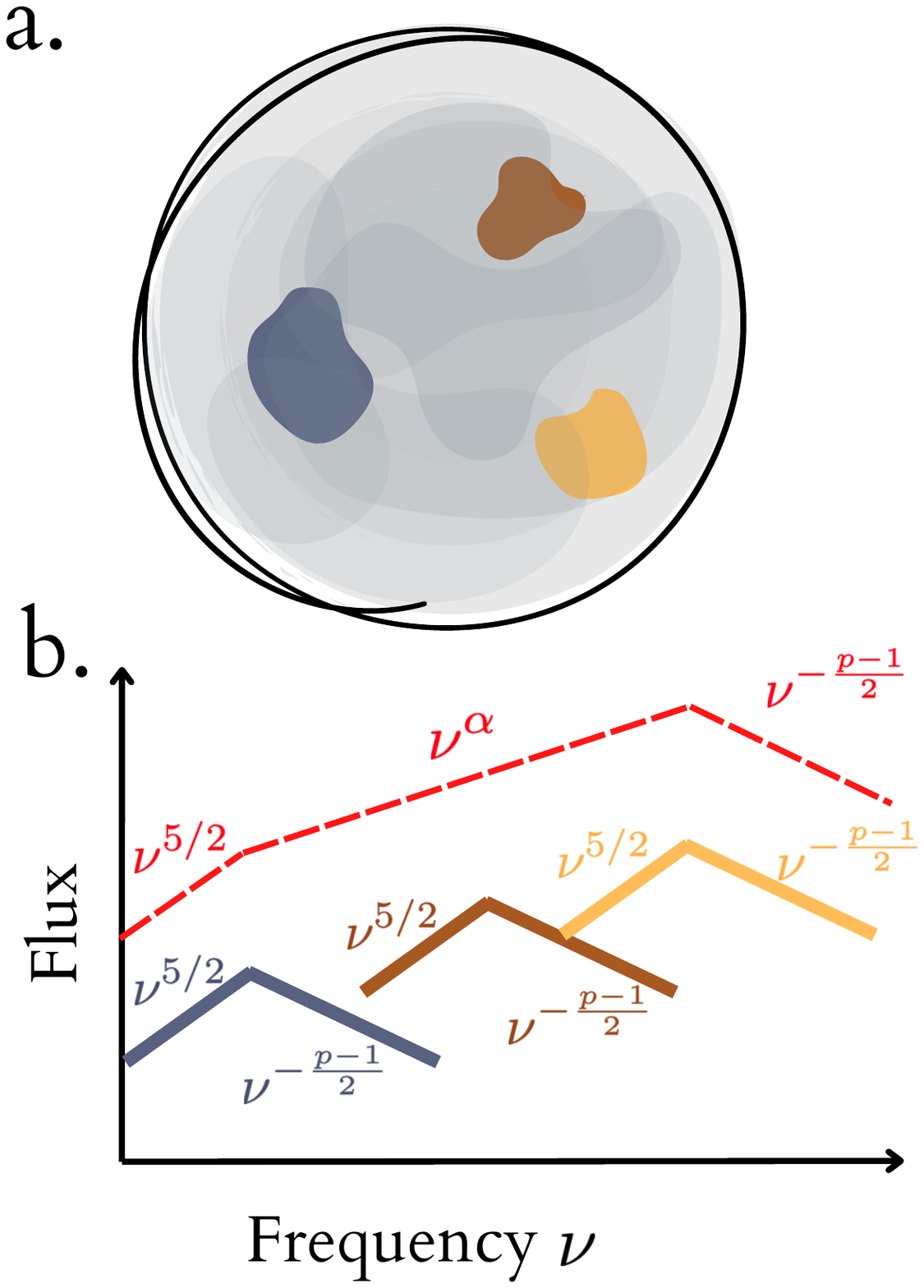}
 \caption{Schematic representation of the model considered in this work. Electrons accelerated by a shock emit synchrotron radiation. In a completely general case, at a fixed time, different regions of the shock front (whose projections onto the plane of the sky are indicated as blue, yellow, and brown patches in panel a), emit spectra with different peak frequencies and amplitudes (blue, yellow, and brown lines in panel b). The observed spectrum (red dashed line) will be the convolution of the emission from the different patches and will show a shallower spectrum than the one expected theoretically.}
  \label{fig:cartoon}
\end{figure}

The problem considered here is illustrated in Figure~\ref{fig:cartoon}. We assume that different patches of the expanding shock front have different magnetic field intensities $B$ or are located at different shock positions $R$ at a given time $t$, i.e., the profile of the curve is due to asymmetries
in the ambient medium and/or to asymmetries in the energy/velocity of the ejecta (see Figure~\ref{fig:cartoon}, panel a). Then, the total spectrum will be obtained by considering the contribution of different emitting regions (each emitting a standard SSA spectrum). When all contributions are added together, a different power-law component can appear in the spectrum (see the $\nu^{\alpha}$ region of panel b in Figure~\ref{fig:cartoon}). The conditions under which this occurs are discussed in detail in this section and the following one.

In the case of a patch moving with non-relativistic velocity, synchrotron emission leads to a characteristic synchrotron self-absorbed spectrum, with 
\begin{eqnarray} 
F_\nu = 1.58 \; F_{\nu_{\rm sa}}  \left(\frac{\nu}{\nu_{\rm sa}}\right)^{5/2} \left[1-e^{-\left(\frac{\nu}{\nu_{\rm sa}}\right)^{-(p+4)/2}}\right]= F_{\nu_{\rm sa}} f_\nu(\nu_{\rm sa})\;,
\label{eq:sinc}
\end{eqnarray}
where $\nu_{\rm sa}$ is the self-absorption frequency, and $F_{\nu_{\rm sa}}$ is the corresponding flux\footnote{We note that $\nu_{\rm sa}$ does not correspond exactly to the peak frequency of the spectrum, nor does $F_{\nu_{\rm sa}}$ correspond to the peak flux. In fact, from Equation 
\ref{eq:sinc} we obtain $\nu_{\rm peak} \simeq 1.1\; \nu_{\rm sa}$ for $p=3$ and $\nu_{\rm peak} \simeq 1.4\; \nu_{\rm sa}$ for $p=2$.} for $\nu = \nu_{\rm sa}$.  The resulting spectrum depends only on the values of $\nu_{\rm sa}$, $F_{\nu_{\rm sa}}$, and $p$. The equation also  shows explicitly that the spectrum can be written as the product of a normalization factor, $F_{\nu_{\rm sa}}$, and a function that depends only on $\nu/\nu_{\rm sa}$ (once the value of $p$ is fixed).

The flux emitted by a single patch in the optically thick part of the spectrum scales as \citep{chevalier98}
\begin{eqnarray}
    F_\nu \propto \eta R^2 B^{-1/2} \nu^{5/2}\;,
    \label{eq:thick}
\end{eqnarray} 
while in the optically thin part scales as 
\begin{eqnarray}
    F_\nu \propto \eta f R^3 B^{(p+5)/2} \nu^{-(p-1)/2}\;,
    \label{eq:thin}
\end{eqnarray} 
where $f = V_{\rm em}/V$ (being $V_{\rm em}$ and $V$ the emitting and the total volume respectively) is the volume filling factor; $f$ describes the radial extension of the emitting region relative to a single value of $B$ and $R$.  $R$ is the shock radius, $B$ is the magnetic field intensity in the emitting region, and $\eta$ is the fraction of the total emitting region covered by a patch located at the shock radius $R$; $\eta$ defines the extent of the emitting region in the plane of the sky, again, relative to a single value of $B$ and $R$. 

The self-absorption frequency can be determined by considering the frequency where equations \ref{eq:thick} and \ref{eq:thin} are identical. We get 
\begin{eqnarray}
    \nu_{\rm sa}\propto f^{2/(p+4)} R^{2/(p+4)}  B^{(p+6)/(p+4)}\;,
    \label{eq:nusa}
\end{eqnarray}
corresponding to a peak flux
\begin{eqnarray}
  F_{\nu_{\rm sa}} \propto \eta f^{5/(p+4)}(R  B)^{(2p+13)/(p+4)}\;.
    \label{eq:Fsa}
\end{eqnarray}
Assuming $p=3$, a value typically observed in radio supernovae, we get $\nu_{\rm sa}\propto (fR)^{2/7}  B^{9/7}$ and $F_{\nu_{\rm sa}} \propto  f^{5/7} (R  B)^{19/7}$.

These equations have been determined by assuming that a fraction $\epsilon_B$ of the post-shock thermal energy $e$ goes into magnetic field energy density, i.e. $\epsilon_B = B^2/(8 \pi e)$. The post-shock thermal energy can be determined using the Rankine-Hugoniot shock-jump conditions. We get $B \propto e^{1/2} \propto \rho^{1/2} v_{\rm sh}$, where $\rho$ is the density of the circumstellar medium and $v_{\rm sh}$ the velocity of the shock front interacting with the circumstellar medium. In the case of a medium shaped by the wind of the progenitor star, $\rho\propto \dot{M}_w/(R^2 v_w)$, being $\dot{M}_w$ the wind mass-loss rate and $v_w$ the wind velocity, yielding $B\propto \dot{M}_w^{1/2} R^{-1} v_w^{-1/2} v_{\rm sh}$. $\nu_{\rm sa}$ and $F_{\nu_{\rm sa}}$, together with the value of $p$, fully determine the shape and normalization of the observed spectra, and depend on the position of the shock and on the density of the environment, shaped by winds/outflows ejected from the progenitor star before the stellar explosion.
As the spectrum depends mainly on the CSM density $\rho$ and on the ejecta velocity $v_{\rm ej}$, it is often complicated to infer whether a certain behaviour in the spectrum originates from properties of the ejecta or from properties of the ambient medium. 

We note that, to derive Equations~\ref{eq:nusa} and \ref{eq:Fsa}, we have followed \citet{chevalier98} by taking the minimum energy of the accelerated electrons to be $E_{\rm min} = m_e c^2$. These equations depend on this assumption, or on any other conditions used for the emission process. However, we will show next that the inversion process can be maintained independently on the physics of the system, and depends only on observation quantities. 
Also, we note that $\eta$ is the generalization of the covering factor $f_B$ defined by \citet{bjornsson2013, bjornsson2017} as the fraction of the emitting region corresponding to a certain value of the magnetic field. We have $f_B = \eta$ in the case of a constant emitting surface $R$ (see Equation~\ref{eq:nusa}).

\section{Inferring the structure of the emitting region}
\label{sec:structure}

In general, the distribution of different emitting patches may be arbitrary. A full mathematical description of the problem requires specifying the fraction of the observed region emitting a spectrum with a certain value of $\nu_{\rm sa}$ and $F_{\nu_{\rm sa}}$, i.e., the two-dimensional distribution $n(\nu_{\rm sa}, F_{\nu_{\rm sa}})$ or the distribution of the magnetic field, shock radius and patch sizes in the emitting region, i.e. $n(B, R, \eta)$. The two formulations are equivalent because $\nu_{\rm sa}$ and $F_{\nu_{\rm sa}}$ can be obtained from $B$, $R$ and $\eta$ following Equations \ref{eq:nusa} and \ref{eq:Fsa}. The distribution $n(\nu_{\rm sa}, F_{\nu_{\rm sa}})$ is formally defined as the fraction of the total emitting region emitting a spectrum with self-absorption frequency ($\nu_{\rm sa}, \nu_{\rm sa}+d\nu_{\rm sa}$) and peak flux ($F_{\nu_{\rm sa}}, F_{\nu_{\rm sa}}+dF_{\nu_{\rm sa}}$).  

By definition, $n(\nu_{\rm sa}, F_{\nu_{\rm sa}})$ fulfills the relations
\begin{equation}
    \iint_0^\infty n(\nu_{\rm sa}, F_{\nu_{\rm sa}}) d\nu_{\rm sa} dF_{\nu_{\rm sa}} = 1\;, \qquad n(\nu_{\rm sa}, F_{\nu_{\rm sa}}) \geq 0\;.
    \label{eq:f}
\end{equation}
Once the function $n(\nu_{\rm sa}, F_{\nu_{\rm sa}})$ is given, the ``observed'' spectrum can be computed by adding together the contributions of all patches, i.e., 
\begin{equation}
    F_{\nu, \rm obs} = \iint_0^\infty n(\nu_{\rm sa}, F_{\nu_{\rm sa}}) F_{\nu}(\nu_{\rm sa}, F_{\nu_{\rm sa}}) dF_{\nu_{\rm sa}} d\nu_{\rm sa}\;,
    \label{eq:Fnu}
\end{equation}    

The simple case of a uniform, symmetric shock corresponds, in this framework, to a density distribution $n(\nu_{\rm sa}, F_{\nu_{\rm sa}})=\delta(\nu_{\rm sa}-\nu_{\rm sa,0}) \delta(F_{\nu_{\rm sa}}-F_{\nu_{\rm sa,0}})$, being $\delta(x)$ the Dirac delta distribution.
In Equation~\ref{eq:Fnu}, $F_{\nu, \rm obs}$ is provided by observations, and $F_\nu(\nu_{\rm sa}, F_{\nu_{\rm sa}})$ is specified by standard synchrotron theory (see Equation \ref{eq:sinc}).
Recovering $n(\nu_{\rm sa}, F_{\nu_{\rm sa}})$ would provide a detailed view of the emitting region, even when the source is unresolved.

Unfortunately, it is not possible to recover $n(\nu_{\rm sa}, F_{\nu_{\rm sa}})$, due to a degeneracy associated with the values of $F_{\nu_{\rm sa}}$. This can be clearly seen by considering different spectra corresponding to the same value of $\nu_{\rm sa}$ and different values of $F_{\nu_{\rm sa}}$. These curves will have an identical shape and the same peak frequency. Therefore, it will be impossible to reconstruct their distribution given the sum of all individual components. For example, a patch emitting a spectrum with $F_{\nu_{\rm sa}}=1$ (in arbitrary units) will be indistinguishable from two patches with $F_{\nu_{\rm sa}}=1/2$. 
Due to the equivalence between this probability distribution and the one defined as a function of $B$ and $R$, this also implies that is not possible to find the distributions of $B$ and $R$ directly from observations without using any extra constraints (for instance, high-resolution images of the emitting source).

\begin{figure*}
\centering
\includegraphics[width=1\textwidth]{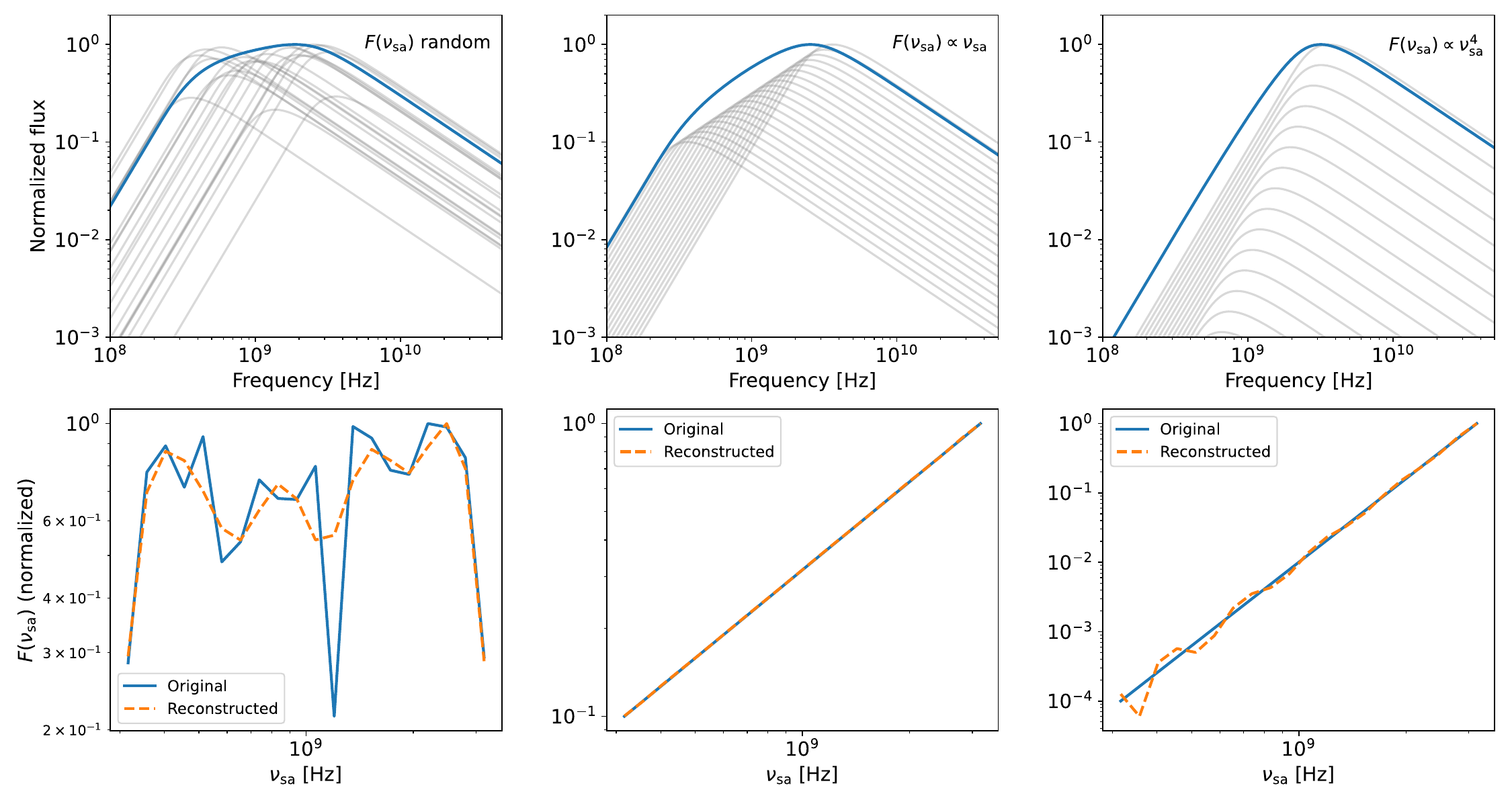}
 \caption{Possible outcomes of the inversion process. In the case of uncorrelated patches (left panel), e.g., with a random variation in the value of $F(\nu_{\rm sa})$ for a given value of $\nu_{\rm sa}$, the contribution of single patches (grey lines, top left panel) adds together, producing a total spectrum (blue line) that preserves the original power-laws. A spread of the flux around the peak is present in this case. In the case of correlated patches, the patches can add together producing a change in the slope of the spectrum (central panels, corresponding to $F(\nu_{\rm sa}) \propto \nu_{\rm sa}$) or not (right panels, corresponding to $F(\nu_{\rm sa})\propto \nu_{\rm sa}^4$), depending on how strong the correlation is (see the text for more details). The bottom panels show the original (blue solid line) and reconstructed (orange dashed line) relations between $F(\nu_{\rm sa})$ and $\nu_{\rm sa}$.}
\label{fig2:tik}
\end{figure*}

\subsection{Model independent, inverse method approach}
\label{model:general}

In this section, we present a method to solve the inverse problem to get information on the different emitting patches.

The spectrum emitted by each patch can be separated into the product of independent functions of $\nu_{\rm sa}$ and $F_{\nu_{\rm sa}}$ (see Equation~\ref{eq:sinc}). Then, Equation~\ref{eq:Fnu} can be rewritten as
\begin{eqnarray}
    F_{\nu, \rm obs} = \iint_0^\infty n(\nu_{\rm sa}, F_{\nu_{\rm sa}}) F_{\nu_{\rm sa}}  dF_{\nu_{\rm sa}} f_\nu(\nu_{\rm sa})d\nu_{\rm sa}\;.
    \label{eq:intf}
\end{eqnarray}
This Equation is still degenerate, as discussed above, but we can remove the degeneracy by defining
\begin{eqnarray}
    F(\nu_{\rm sa})= \int_0^\infty n(\nu_{\rm sa}, F_{\nu_{\rm sa}}) F_{\nu_{\rm sa}}
    dF_{\nu_{\rm sa}}\;,
\end{eqnarray}
where $F(\nu_{\rm sa})$ is the peak flux obtained by adding together all patches with the same value of $\nu_{\rm sa}$, i.e. by considering the emitting surface characterised by a single value of the self-absorption frequency $\nu_{\rm sa}$ and peak flux $F_{\nu_{\rm sa}}$. This patch can be composed of several disconnected emitting regions, all characterised by the same value of $\nu_{\rm sa}$. 

The observed flux is then related to the emission of each patch as:
\begin{eqnarray}
    F_{\nu, \rm obs}= \int_0^\infty F(\nu_{\rm sa}) f_\nu(\nu_{\rm sa})
    d\nu_{\rm sa}\;.
\label{eq:inv}
\end{eqnarray}
The simple case of a uniform, symmetric shock corresponds, in this framework, to a peak flux $F(\nu_{\rm sa})=\delta(F_{\nu_{\rm sa}}-F_{\nu_{\rm sa,0}})$. In this case, the integral in Equation \ref{eq:inv} reduces to Equation \ref{eq:sinc}.

By inverting Equation~\ref{eq:inv}, we can recover $F(\nu_{\rm sa})$ as a function of $\nu_{\rm sa}$. In the framework described in this paper, $F(\nu_{\rm sa})$ is then the fundamental quantity, characterising the emitting region. 

Equation \ref{eq:inv} belongs to the class of so-called inverse problems, and it is ill-posed (i.e. the solution is non-unique in general and neither is it stable if noise is present in the data). Then, Tikhonov \citep[see, e.g.,][]{aster2018} regularization techniques must be employed to solve it. Equation \ref{eq:inv} is inverted by solving the Equation
\begin{eqnarray}
\min (||Ax - B||^2 + \alpha^2 ||Lx||^2)\;,
\label{tik}
\end{eqnarray}
where the term $\alpha||Lx||^2$ penalizes high-frequency fluctuations in the solution, being $\alpha$ a regularisation parameter. Formally, $||Lx||^2 = \sum_i (x_{i+1}-x_i)^2$. 

Figure~\ref{fig2:tik} illustrates the possible outcomes of this inversion process. 
The blue full curves in the bottom panels show the function $F(\nu_{\rm sa})$ chosen in each case considered. From left to right, these correspond to a random variation in the values of $F(\nu_{\rm sa})$, $F(\nu_{\rm sa}) \propto \nu_{\rm sa}$ and $F(\nu_{\rm sa}) \propto \nu_{\rm sa}^4$ respectively. $\nu_{\rm sa}$ and $F(\nu_{\rm sa})$ are uncorrelated in the left panel, and correlated in the central and right panels. 

The grey lines shown in the top panels of Figure~\ref{fig2:tik} show the SSA spectrum, obtained by sampling the values of $F(\nu_{\rm sa})$ and $\nu_{\rm sa}$, shown in the bottom panel, over 20 patches. The blue line shown in the top panels represents the total emission (normalized to the peak value) obtained by adding the contributions of each single patch. Finally, the orange lines in the bottom panels represent the 
$F({\nu_{\rm sa}})$ vs $\nu_{\rm sa}$ relation obtained inverting Equation~\ref{eq:inv}.

In the case of uncorrelated patches, each value of $\nu_{\rm sa}$ corresponds to a random variation of $F({\nu_{\rm sa}})$. Then, adding the contributions from the different patches, the spectrum preserves the same spectral slopes seen in a standard SSA spectrum. In this case, the optically thick slope remains equal to 5/2, while the spread into the values of $\nu_{\rm sa}$ directly reflects in a broad transition region between the two power-law segments. This behaviour is sometimes observed in radio supernovae \citep[for instance, SN 2023ixf, see][]{nayana2025b}.

We note that, to estimate how broad a curve is, the spectrum can be fitted adopting a smoothed broken power-law of the form 
\begin{eqnarray}
F_\nu = F_{\nu_{\rm sa}}  \left[ \left(\frac{\nu}{\nu_{\rm sa}}\right)^{-s\alpha_1} + \left(\frac{\nu}{\nu_{\rm sa}}\right)^{-s \alpha_2}\right]^{-1/s}\;,
\label{eq:fit}
\end{eqnarray}
where $\alpha_1 = 5/2$ and $\alpha_2=-(p-1)/2$ for an homogeneous source, but can be different in general.
The value of the parameter $s$ controls how smooth is the transition between the optically thick and thin portions of the spectrum obtained, and can be compared with theoretical predictions.
This approach has been followed recently to fit spectra of SN 2012ap \citep{sfaradi2026} and SN 2012au (Wiston et al., in preparation).

The central and right panels of Figure~\ref{fig2:tik} show the cases of a strong correlation between $\nu_{\rm sa}$ and $F({\nu_{\rm sa}})$, i.e. in which an increase of $\nu_{\rm sa}$ corresponds to an increase in $F({\nu_{\rm sa}})$. In both cases, the transition between low- and high- frequencies is broader than in the homogeneous case (see \citealt{bjornsson2013, bjornsson2017}). 
The central and right, top panels also show that having a strong correlation between $\nu_{\rm sa}$ and $F({\nu_{\rm sa}})$ is not a sufficient condition to change the low-frequency slope. In the examples shown in Figure \ref{fig2:tik}, in fact, the low-frequency slope changes when $F({\nu_{\rm sa}}) \propto \nu_{\rm sa}$, but remains unchanged (and $\propto \nu^{5/2}$) when $F({\nu_{\rm sa}}) \propto \nu_{\rm sa}^4$.

Finally, we notice that the inversion process recovers accurately the original patch distribution in the case of correlation among the patches  (central and right, bottom panels), and provides information on the structure of the emitting region also in the case of uncorrelated emitting patches (left, bottom panels).

To see how the degree of correlation between $F(\nu_{\rm sa})$ and $\nu_{\rm sa}$ determines the resulting spectrum, we consider 
a distribution of peak fluxes given by
$F(\nu_{\rm sa}) = k \nu_{\rm sa}^\alpha$ for $\nu_{\rm sa,min} < \nu_{\rm sa} < \nu_{\rm sa,max}$,
and $F(\nu_{\rm sa})=0$ otherwise. 
Equation \ref{eq:inv} leads to
\begin{eqnarray}
    F_\nu = \int_{\nu_{\rm sa,min}}^{\nu_{\rm sa,max}} k \nu_{\rm sa}^\alpha \left(\frac{\nu}{\nu_{\rm sa}} \right)^{5/2} \left[1-e^{-(\nu/\nu_{\rm sa})^{-(p+4)/2}} \right]
    d\nu_{\rm sa}\;,
\end{eqnarray}
Setting $x=\nu/\nu_{\rm sa}$, we get
\begin{eqnarray}
    F_\nu = k \nu^{\alpha+1} \int_{\nu/\nu_{\rm sa,max}}^{\nu/\nu_{\rm sa,min}}x^{1/2-\alpha} \left[1-e^{-x^{-(p+4)/2}} \right]
    dx\;,
\end{eqnarray}
For $\nu < \nu_{\rm sa,min}$, we get $F_\nu \propto \nu^{5/2}$, 
and for $\nu > \nu_{\rm sa,max}$, $F_\nu \propto \nu^{-(p-1)/2}$. For intermediate frequencies, the integral can be easily evaluated for
$\nu_{\rm sa,min} \ll \nu \ll \nu_{\rm sa,max}$ (whence the integration limits extend from 0 to $\infty$), to give
\begin{equation}
F_\nu \propto 
\begin{cases} 
\nu^{-(p-1)/2} & \text{if } \qquad \alpha < -(p+1)/2 \\[2mm]
\nu^{1+\alpha} & \text{if }  \qquad -(p+1)/2 < \alpha < 3/2 \\[1mm]
\nu^{5/2} & \text{if }  \qquad \alpha > 3/2
\label{eq:alpha}
\end{cases}
\end{equation}%

\begin{figure}
\centering
\includegraphics[width=.45\textwidth]{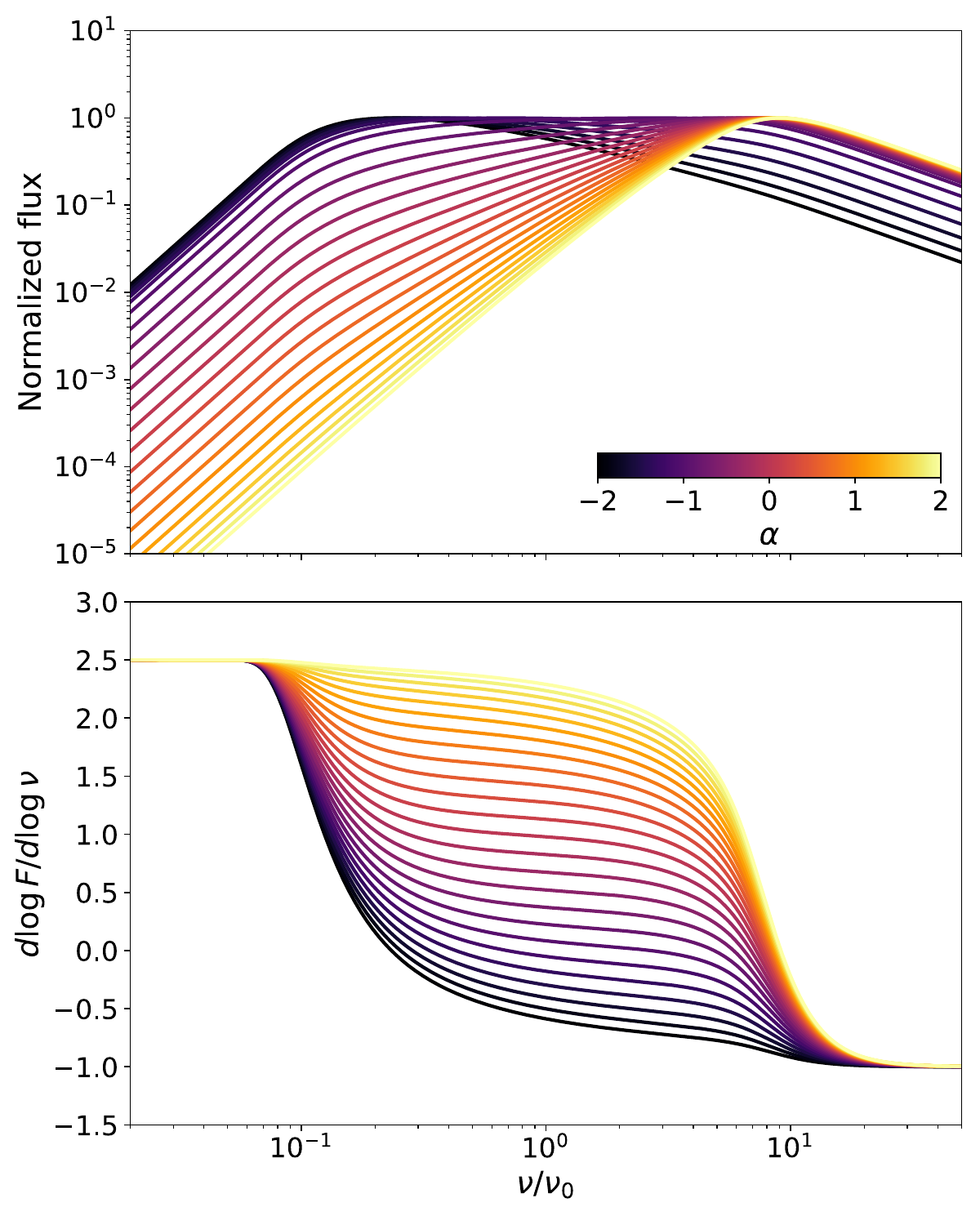}
 \caption{
 Spectra computed for different values of $\alpha$, relating the peak flux to the self-absorption frequency $F(\nu_{\rm sa}) = k \nu_{\rm sa}^\alpha$. The top panel shows the total flux, obtained by adding the contribution of the 200 patches sampled in the range $\nu_{\rm sa} = 0.1-10$, while the bottom panel shows the resulting spectral slope.}
  \label{fig:alpha}
\end{figure}

Figure~\ref{fig:alpha} illustrates the spectra obtained by adding individual patches (upper panel), and the corresponding spectral slopes (bottom panel). The different patches are defined considering $F(\nu_{\rm sa}) = k \nu_{\rm sa}^\alpha$, and adding the contribution of 200 patches, centred at $\nu=\nu_0$ and sampled in the range $(\nu_{\rm sa_{\rm min}}, \nu_{\rm sa_{\rm max}})$ = (0.1,10). 
For $\nu/\nu_0 < 0.1$, the spectral slope is $5/2$, while for $\nu/\nu_0 > 10$, the spectral slope is  $-(p-1)/2$ (we employ $p=3$). For positive values of $\alpha$, the peak corresponds to $\nu/\nu_0 = 10$, and the slope in the range $ 0.1< \nu/\nu_0 < 10$ increases for smaller slopes of $\alpha$. These cases correspond to observations with low-frequency (optically thick) deviations from the standard $5/2$ slope. 
For negative values of $\alpha$, the peak corresponds to $\nu/\nu_0 = 0.1$, and the slope in the range $ 0.1< \nu/\nu_0 < 10$ drops as $\alpha$ becomes more negative. 
These cases correspond to observations with high-frequency (optically thin) deviations from the standard $-(p-1)/2$ slope. 
In the case of $\alpha = -1$, the spectrum is flat. This has been discussed for the first time in the seminal paper by \citet{blandford1979} to explain observations of AGNs.
These results are completely consistent with Equation  \ref{eq:alpha}.

We conclude that for the low-frequency spectral slope to be modified by the emission from different patches, the peak flux and self-absorption frequencies must be correlated, but not too strongly (i.e., with a dependence shallower than $\nu_{\rm sa}^{3/2}$). Steeper dependencies can produce a broadening of the transition between the low- and high-frequency parts of the spectrum, but leave the low-frequency spectral slope unchanged.
Then, a low-frequency spectrum $F_\nu \propto \nu^{5/2}$ can be produced either by an homogeneous source, or by an inhomogeneous source with an effective flux scaling as $F(\nu_{\rm sa}) \propto \nu_{\rm sa}^{\alpha}$, with $\alpha>3/2$. The case of an homogeneous emitting region is recovered when $\alpha\rightarrow \infty$.

\subsection{Implications on the physical structure of the emitting region}
\label{model:mot}

In the previous section, we discussed the role of inhomogeneity in shaping the SSA spectrum. Here, we discuss the implications on the physical structure of the emitting region. 
For this purpose, we consider the \citet{chevalier98} model, presented in section \ref{sec:methods}. For simplicity, we consider the simple power-law relation $F(\nu_{\rm sa}) = k \nu_{\rm sa}^\alpha$ between the self-absorption frequency and the peak flux corresponding to each emitting patch, to get a general insight into the problem, but we remind that $F(\nu_{\rm sa})$ can be a complex function of $\nu_{\rm sa}$ in general. The exponent $\alpha$ is related to the observed spectral slope by Equation \ref{eq:alpha}.

From Equations~\ref{eq:nusa} and \ref{eq:Fsa}, we get a covering factor
\begin{eqnarray}
    \eta \propto \left(f^{2\alpha-5} R^{2 \alpha-2p-13} B^{\alpha (p+6)-2p-13}\right)^{1/(p+4)}\;.
    \label{eq:eta}
\end{eqnarray}
In the following, we will take $f$ to be constant. This is something commonly assumed in studies of radio supernovae, and it is justified by the fact that the size of the post-shock region, where the synchrotron emission originates, is not expected to vary substantially, even in presence of density fluctuations in the  circumstellar medium. In addition,  variations in $f$ have a weaker effect on $\eta$ than variations in $R$ and $B$. This can be seen directly from equation \ref{eq:eta}. For simplicity, we take $p=3$. Then, we have $\eta \propto (fRB)^{(2\alpha-5)/7} R^{-2} B^{\alpha-2}$, which implies that $\eta$ is more sensitive to variations in $R$ and $B$ than $f$ (for instance, for $\alpha=1$ we get $\eta \propto R^{-2.4}B^{-1.4}f^{-0.4}$). 
Finally, we note that the value of  $f$ can be determined through numerical simulations, where the size of the emitting region, and its dependence on the other parameters, can be resolved directly. This is left for future work.

We first discuss the case $R=$ constant. Again, by taking 
$p=3$, Equation~\ref{eq:eta}
reduces to $\eta \propto B^{(9\alpha - 19)/7}$.
In this case, $\eta$ increases with the magnetic field, if $\alpha > 19/9$, and drops with the magnetic field for $\alpha < 19/9$.
As $\nu_{\rm sa} \propto B^{9/7}$, the former case corresponds to patches with larger magnetic field occupying a larger fraction of the emitting surface, while in the latter case the larger emitting surface corresponds to smaller magnetic fields.

The general case is much more complex, and depends on the precise value of $\alpha$. If we observe a spectrum with a slope $\alpha_{\rm obs}$, Equation~\ref{eq:alpha} provides the corresponding value of $\alpha= \alpha_{\rm obs}-1$. 
Then, the implications, in terms of magnetic field and position of the emitting regions, are the following:

\begin{itemize}
    \item {$\alpha < -1$}: 
     the spectrum drops as $\nu^{\alpha+1}$ above the self-absorption frequency (Equation~\ref{eq:alpha}).
The emission close to the observed  self-absorption frequency ($\nu_{\rm sa} \sim \nu_{\rm sa, min}$ - see figure   \ref{fig:alpha}) is dominated by patches with small values of $\nu_{\rm sa}$. $\eta$ increases for small values of $R$ and $B$ (Equation~\ref{eq:eta}). Then, patches with small values of radius and magnetic field dominate the emission in this case, as they occupy a very large fraction of the total emission region.
    \item {$-1 < \alpha < 3/2$}:  in this case, the spectrum increases as $\nu^{\alpha+1}$ below the self-absorption frequency 
    (Equation~\ref{eq:alpha}). 
The emission close to the observed  self-absorption frequency is dominated by patches with large values of $\nu_{\rm sa}$, and $\eta$ increases for small values of $R$ and $B$ (Equation~\ref{eq:eta}) as in the previous case. The change in the spectral slope is due to the contribution of patches with moderately small magnetic fields and small radii.
 \item {$3/2 < \alpha < 19/9$}: 
$\eta$ still increases for small values of $R$ and $B$ (Equation~\ref{eq:eta}). The slope of the low-frequency spectrum is very close to $5/2$. Nevertheless, the different patches still contribute to the broadening of the spectrum around $\nu_{\rm sa}$.
    \item {$19/9 < \alpha$}: 
In this case, $\eta$ increases either for small values of $R$ (for $\alpha > 19/2$), or large values of $R$ (for $\alpha > 19/9$),   
    and also increases with larger values of $B$ (Equation~\ref{eq:eta}). As the value of $\alpha$ increases, larger portions of the emitting regions are occupied by regions with large magnetic field, dominating the emission and obscuring regions with smaller magnetic fields.
\end{itemize}

\citet{bjornsson2013, bjornsson2017} discussed how the estimation of the shock radius, when the emitting region is taken as homogeneous, leads to an underestimation of the true shock radius, once possible inhomogeneities are considered. 
As a consequence, estimates of the shock radius determined from  unresolved radio observations can be lower than the shock radius determined by VLBI observations, if inhomogeneities in the emitting region are present. A comparison between shock radii estimated from SSA and from VLBI observations has been presented by \citet{chandra2024a} for SN 1993J, although the two estimates yielded similar results in that case.

In the following, we show how, with the formalism developed in this paper, it is possible to estimate the true shock radius, given the value of the radius computed using SSA. The shock radius depends directly on the peak flux (see Equation~\ref{eq:Fsa}). When multiple patches are present, the peak flux is spread over a certain range of self-absorption frequencies, while only one self-absorption frequency is present in the case of a homogeneous source. As a consequence, the flux in the inhomogeneous case is lower, which leads to underestimate the shock radius. 

The correcting factor in the radius can be obtained by estimating the ratio between the peak flux in the inhomogeneous and homogeneous cases. As shown in Appendix~\ref{ap:A}, the ratio depends on the values of $\alpha$, $p$ as well as that for the self-absorption frequency. 

\begin{figure}
\centering
\includegraphics[width=.5\textwidth]{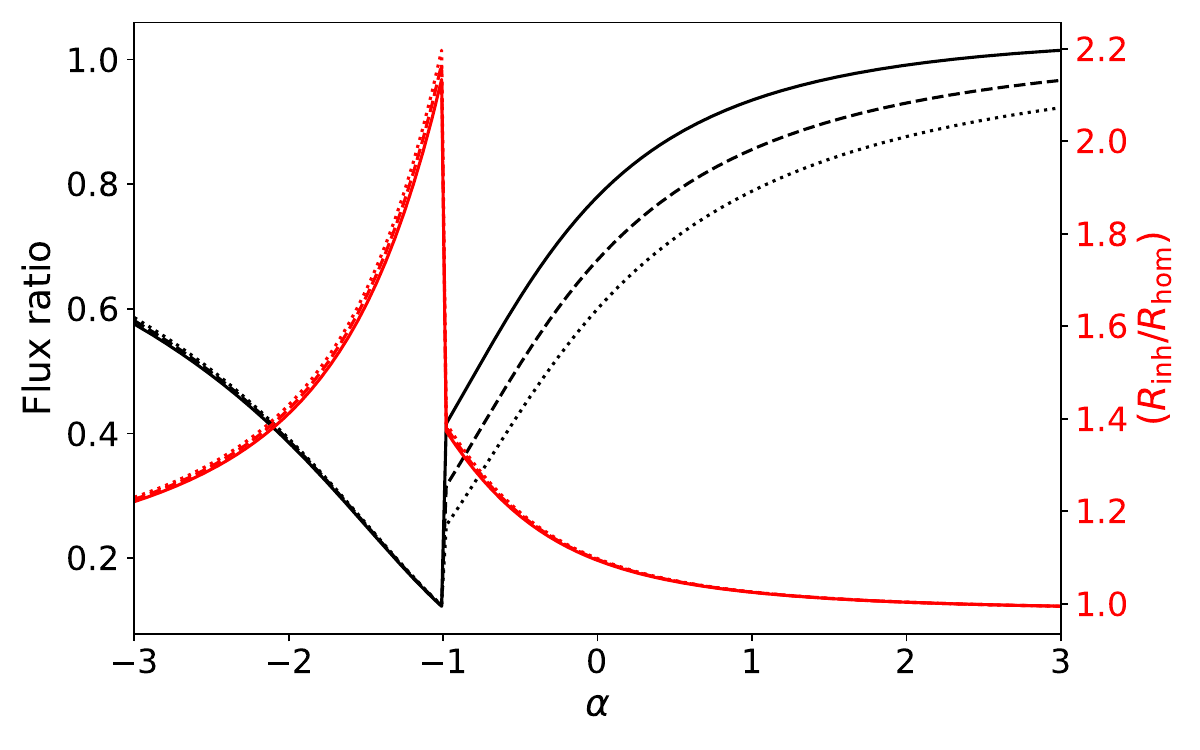}
 \caption{
 The black lines show the ratio of the peak flux emitted by an inhomogeneous source, with respect to the homogeneous counterpart. Red lines correspond to the 
 correction factor to apply to get the true radius. Solid, dashed, and dotted lines correspond to $p=2.5, 3, 3.5$ respectively.}
  \label{fig:flux}
\end{figure}

Figure~\ref{fig:flux} shows the ratio between the peak flux for inhomogeneous to homogeneous sources (black lines), as a function of $\alpha$ and for different values of $p$, and the correction factor to estimate the real value of the radius. For values close to $\alpha=-1$, the effect can be very large, and leads to an underestimation of the true radius (and velocity) by a factor of $\sim 2$. On the other hand, the dependence on $p$ is weak. 

Then, the following strategy can be adopted: from the low-frequency and high-frequency slopes, the values of $\alpha$ and $p$ can be determined. Then, integrating Equation~\ref{eq:ratio}, we can obtain the true radial position and velocity (if we know the radius at different times). We note that, while in Equation~\ref{eq:ratio} the integral depends on the particular form used for $F(\nu_{\rm sa}) = k \nu_{\rm sa}^\alpha$, the same method can be employed to determine the shock velocity once the inverse problem is solved and a general function (i.e. not necessarily a power-law) for $F(\nu_{\rm sa})$ is determined (see Equation~\ref{eq:a6}).

The opposite is also true. When VLBI observations are available, it is possible to infer the degree of inhomogeneity of the emitting region by comparing the radius estimated from VLBI with the shock radius estimated from unresolved SSA observations.

\section{Application to astrophysical sources}
\label{sec:applications}

In this section, we describe how the framework presented in the previous section can be used to infer the degree of asymmetry present in the emitting region across a variety of high-energy astrophysics phenomena. In particular, we present applications to radio supernovae (SNe) and fast blue optical transients (FBOTs). The same techniques can be extended to detect asymmetries in other radio sources such as tidal disruption events or gamma-ray bursts, although in those cases relativistic effects should also be considered.

\subsection{Inhomogeneities in young supernovae}

Supernovae are expected to be associated with some degree of asymmetry. Asymmetries in the ejecta are produced in three-dimensional numerical simulations of neutrino-driven supernova explosions \citep[see, e.g., ][and references therein]{Mezzacappa2020}, and in central engine driven supernova explosions, as SN associated with rapidly rotating magnetars \citep[e.g.,][]{Bucciantini2009, Metzger2015}, collapsars \citep{1993ApJ...405..273W} from Kerr black holes \citep{2011ApJ...727...29M}, or core-collapse SNe driven by the jittering-jet mechanism \citep[see, e.g.][]{Papish2014, Soker2022}. In some cases, supernovae show evidence of asymmetry in the emitting region. For instance, radio and X-ray observations of SN 2014C \citep{Brethauer22, Vargas22} and SN 2023ixf \citep{nayana2025b} seem to originate from regions expanding with very different velocities.

While optical emission tracks the bulk of the ejecta, which contains most of the ejecta's energy and mass, radio emission traces the forward shock. This region has a higher velocity than the bulk of the ejecta, i.e. $v \sim 0.3$~c in typical core-collapse supernovae, compared to $\sim 10^4$ km s$^{-1}$.
Then, the radio emission from the forward shock
tracks inhomogeneities in the ambient medium, associated with material lost from the progenitor star decades to centuries before the explosion, in addition to inhomogeneities in the ejecta, originated into the explosion mechanism itself.

\begin{figure}
    \centering
\includegraphics[width=\linewidth]{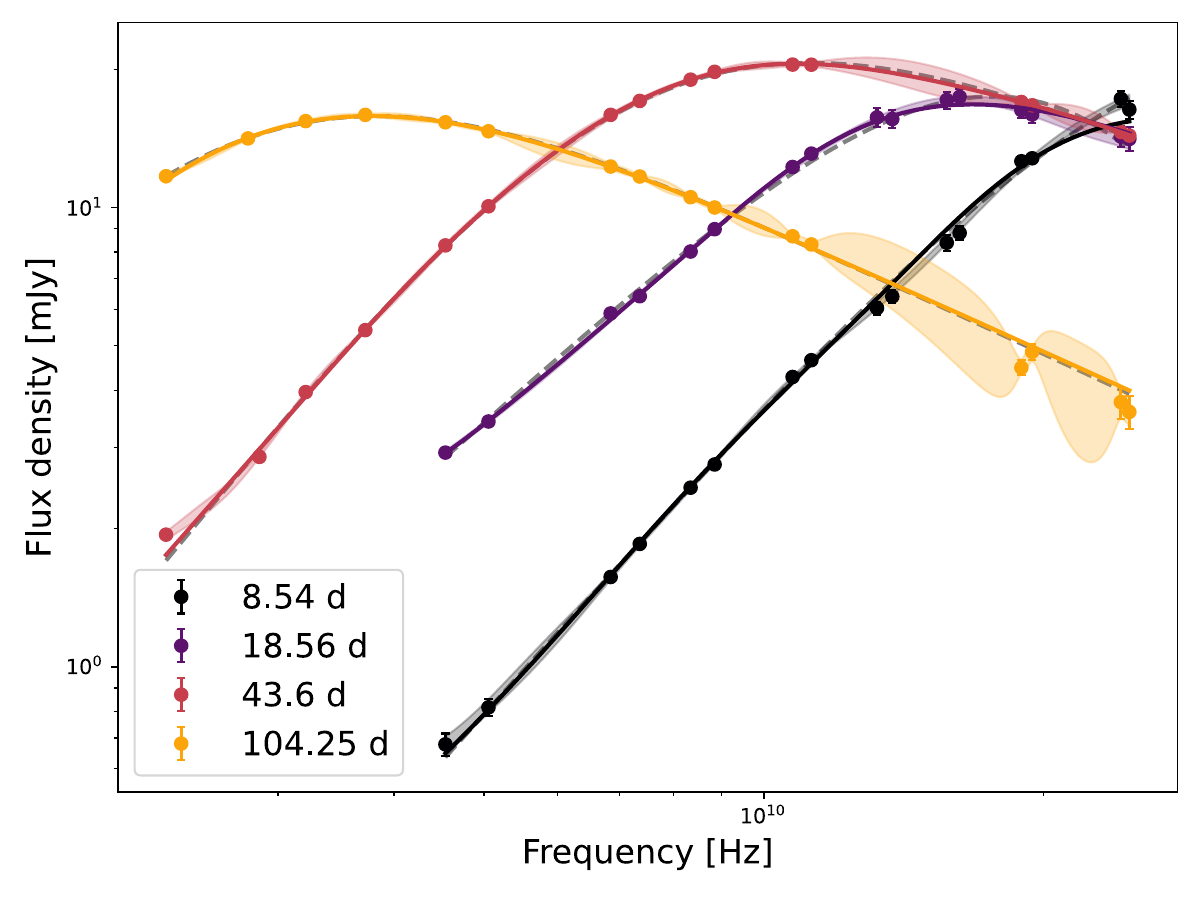}
\caption{Spectrum of SN 2016coi. The black, violet, red, and yellow points correspond to observations at 8.54, 18.56, 43.6, and 104.25~days \citep[][]{nayana2020}. The lightly filled regions show the Gaussian process fits (with 1000 interpolation points per epoch). The dashed line represents the fit obtained using Equation~\ref{eq:fit}. Finally, the solid lines show the results of the multi-patch inversion process.
}
\label{fig:2012ap}
\end{figure}

While hundreds of SNe have been observed in radio \citep[e.g.,][]{bietenholz21}, only a few dozen have been followed extensively, and only a handful of cases have extensive covering below the self-absorption frequency. One of the clearest case is the SN 2016coi \citep{terreram2019, nayana2020}, for which multi-epoch follow-up over a wide range of frequency (see Figure~\ref{fig:2012ap}) is available. In the following, we apply the model discussed in the previous section to this supernova.

\begin{figure}
    \centering
\includegraphics[width=\linewidth]{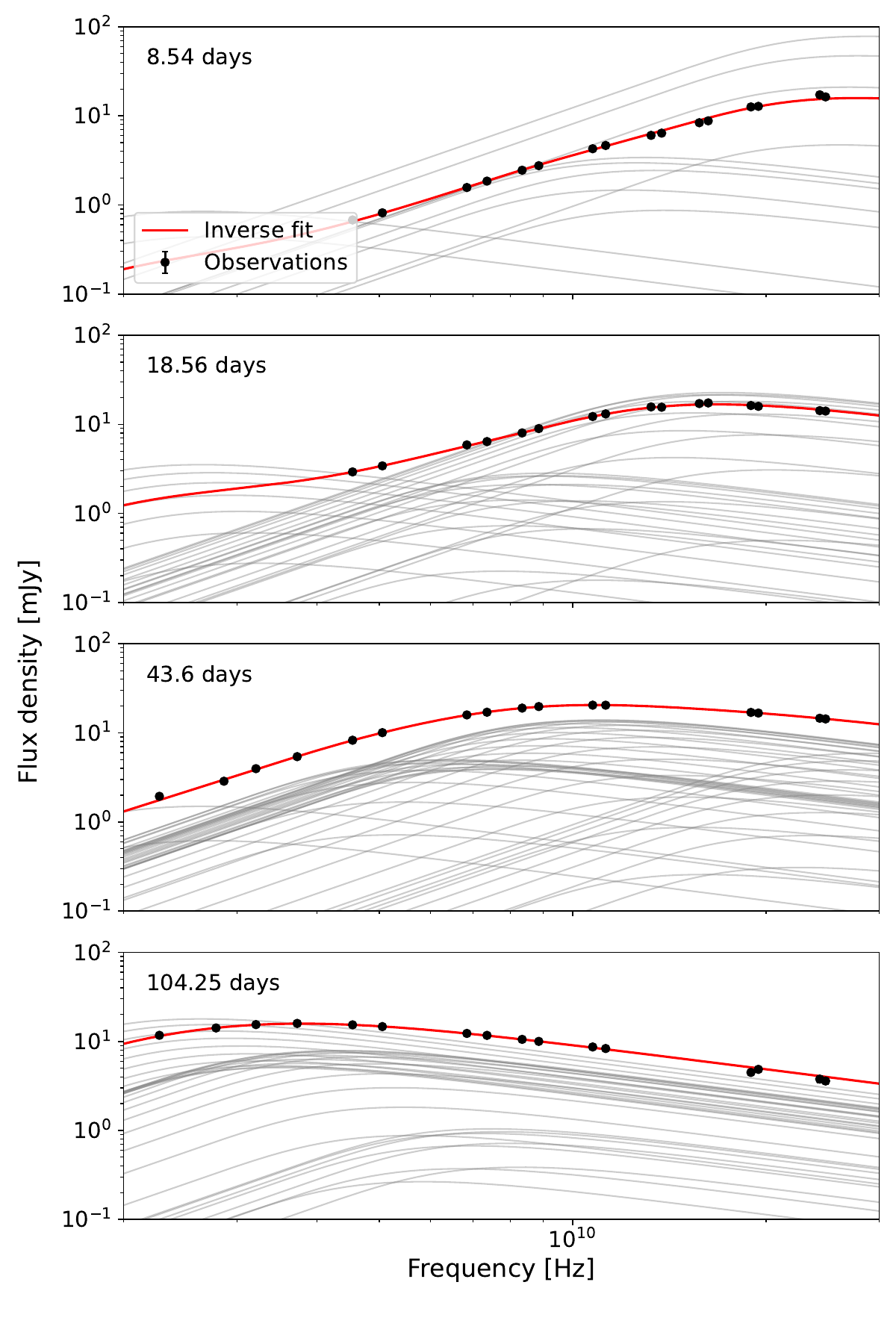}
\caption{Observations and models of SN 2016coi at four different epochs. From top to bottom: 8.54, 18.56, 43.6, and 104.25~days. The black dots represent the observations (error bars are smaller than the marker size). The grey lines represent the contribution of single patches (a total of 80 patches have been used). The solid red line represents the sum of all patches. For clarity, the emission from the individual patches is multiplied by a factor of 10. }
\label{fig:patches}
\end{figure}

\begin{figure}
    \centering
\includegraphics[width=\linewidth]{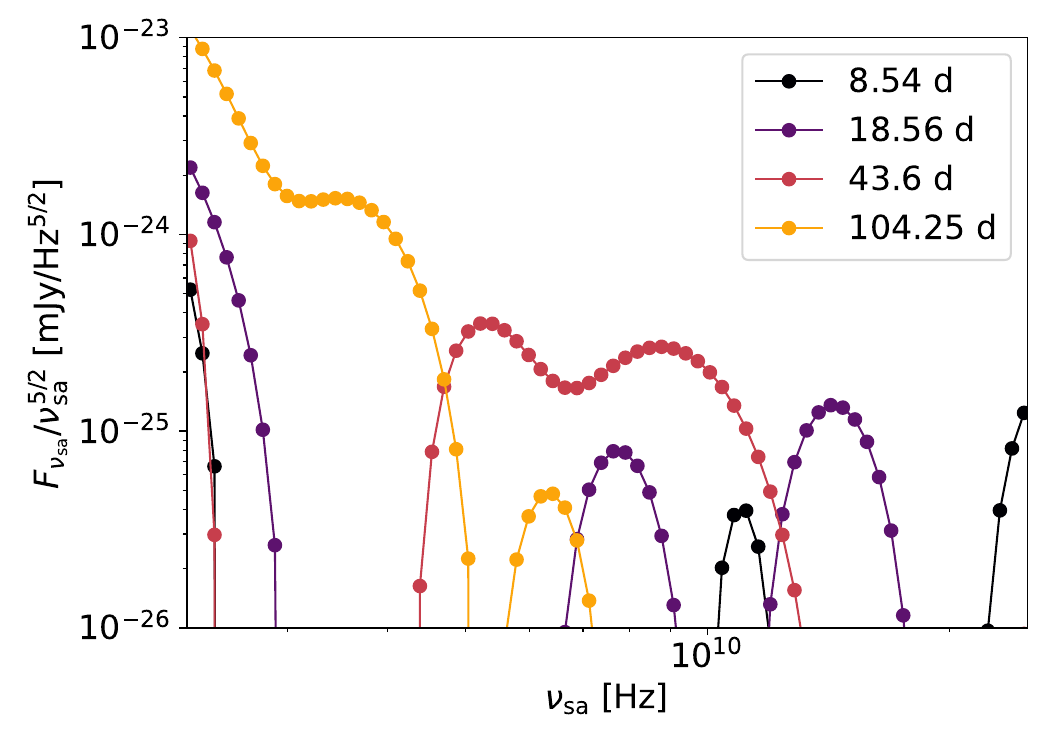}
\caption{
Normalized peak flux $F_{\nu_{\rm sa}}/\nu_{\rm sa}^{{5/2}}$ for the SN 2016coi, as a function of the self-absorption frequency $\nu_{\rm sa}$ for each patch, at four epochs: 8.54~days (black), 18.56~days (violet), 43.6~days (red) and 104.25~days (orange). We use 70 patches, shown as dots in the Figure.}
\label{fig:Fpvsnusa}
\end{figure}

SN 2016coi, also known as ASASSN-16fp, is a broad-line SN~Ic located at a distance of 19.68~Mpc \citep{kumar2018}. Its peak absolute magnitude was $M_V = -17.7 \pm 0.2$ mag, comparable to other broad-line Ic SNe as SN 2002ap and SN 2012ap, but fainter than SNe associated with gamma-ray bursts \citep{kumar2018}.

Radio observations covered a wide range of frequencies, from 0.33 to 25~GHz, and from 8 to 1136~days post-explosion  \citep{terreram2019,nayana2020}. These observations show signatures of an increase in the CSM density at a distance of $\sim 1.1 \times  10^{16}$~cm.

Radio data used in the analysis presented here are shown in Figure~\ref{fig:2012ap}, ranging from 8.54 to 104.25~days. 
To determine the slope $p$ of the accelerate electron population, we fit observations at the four epochs with the function \ref{eq:fit}.
The fits are shown as dashed lines in the Figure.
We get the following values from the fitting process:
\begin{eqnarray}
\alpha_1 &=& 2.267 \pm 0.081 \qquad {\rm at\; 8.54\; days\;,} \\
\alpha_1 &=& 1.845 \pm 0.048\qquad  {\rm at\; 18.56\; days\;,} \\
\alpha_1 &=& 2.534 \pm 0.075 \qquad  {\rm at\; 43.6\; days\;,}\\ 
\alpha_1 &=& 2.202 \pm 0.285 \qquad {\rm at\; 104.25\; days\;.}     
\end{eqnarray}
In the standard, homogeneous SSA model, a slope of $\alpha_1 = 2.5$ is expected. In three out of the four epochs, the slope is flatter than the expected value of 2.5 (at 2.9, 13.6, 1.0 $\sigma$, respectively, during the first, second, and fourth epoch). In addition, the spectrum is very broad close to $\nu_{\rm sa}$ at all epochs.
Thus, we conclude that observations clearly rule out the possibility of a homogenous, single patch emitting region, and are consistent with a multi-patch SSA spectrum.
In addition, the case of free-free absorption is also ruled out, as the slope of the low-frequency part of the spectrum is steeper than the expected slope of 2 characteristic of free-free absorption in all but one epoch.

At high frequencies, we get $\alpha_2 = -0.820 \pm 0.052$ at 43.6 days, and $\alpha_2 = -0.945 \pm 0.026$ at 104.25 days, corresponding to a weighted average value of 
\begin{eqnarray}
p=2.84 \pm 0.05\;,
\label{eq:p}
\end{eqnarray}
taking $\alpha_2 = -(p-1)/2$.

Before inverting Equation~\ref{eq:inv} to find $F(\nu_{\rm sa})$, we interpolate the spectrum using a Gaussian process regression (GPR) approach. This is done using the \emph{GaussianProcessRegressor} class from the \emph{scikit-learn} library, for which a composite kernel from the \emph{sklearn.gaussian\_process.kernels} module was used, allowing both smooth trends and local variations to be modelled. The model was trained by searching for the best hyperparameter values in the kernel by maximizing the marginal log-likelihood of the observed data, incorporating experimental uncertainties as noise. From this, a grid of 1000 points was generated on which the flux values and their standard deviations were predicted. In Figure~\ref{fig:2012ap}, the light-coloured filled region contains $95\%$ of the flux values at each frequency.  

To invert Equation~\ref{eq:inv}, we implement a Tikhonov regularization model, as described by Equation~\ref{tik}.
The Tikhonov method depends on two parameters: the number of patches $N_p$, and the regularization parameter $\alpha$. Increasing $N_p$ adds more free parameters to the model, which leads to better fits and smaller values of the reduced $\chi^2$. In contrast, increasing $\alpha$ produces smoother solutions, but it typically leads to larger values of the reduced $\chi^2$. To choose the values of $N_p$ and $\alpha$, we explore different values of $N_p$ (from 10 to 150). For each choice of $N_p$, we increase $\alpha$ until the reduced $\chi^2 \lesssim 1-1.2$. We found that the reconstructed function $F(\nu_{\rm sa})$ converges from $N_p \gtrsim 50$. Then, we employ 70 patches and $\alpha=140$ for the final reconstruction.

Figure~\ref{fig:patches} shows the result of the inversion process. Grey lines represent the standard SSA spectrum obtained for each patch. Red solid lines represent the sum of the emission from individual patches. The comparison with observations (represented by dots in Figure~\ref{fig:patches}) shows that the agreement is excellent (see also Figure~\ref{fig:2012ap}). Also, Figure~\ref{fig:patches} shows that in all cases several patches are needed to correctly reproduce the observations. On one hand, this is a direct result of using a large value of the regularisation parameter, which tends to create smoother solutions. On the other hand, this is directly related to the slope being flatter than 2.5 in three out of the four epoch. Interestingly, several patches are needed to explain observations also during the third epoch, which has $\alpha_1 = 2.534 \pm 0.075$, consistent with the standard, single shell SSA model. This is related to the broad transition between low- and high-frequencies, which cannot be well modelled by a single homogeneous emitting region.

Figure~\ref{fig:Fpvsnusa} shows the distribution of the peak flux as a function of the self-absorption frequency, for all the reconstructed patches. Since the standard single-patch SSA spectrum scales as $F_\nu \propto \nu^{5/2}$, we show $F_{\nu_{\rm sa}} / \nu_{\rm sa}^{5/2}$ in the Figure, to make it easier to see which patches dominate the flux at each frequency.

The results shown in Figure~\ref{fig:Fpvsnusa} can be interpreted in terms of the physical parameters $\eta$, $B$, $R$, by considering that (see Equations~\ref{eq:nusa} and \ref{eq:Fsa}, or directly Equation~\ref{eq:thick})
\begin{eqnarray}
    \frac{F_{\nu_{\rm sa}}}{\nu_{\rm sa}^{5/2}} \propto \eta R^2 B^{-1}\;,
\end{eqnarray}
and 
\begin{eqnarray}
    \nu_{\rm sa}\propto R^{0.29}  B^{1.29}\;,
    \label{eq:nusasn}
\end{eqnarray}
where we have used Equation~\ref{eq:nusa} and $p=2.84$ (see Equation~\ref{eq:p}).

The first and second epochs (see Figure~\ref{fig:Fpvsnusa}) are dominated by a few patches, peaking around two characteristic frequencies, i.e. $\nu_{\rm sa}\sim 10$ GHz and $\nu_{\rm sa}\gtrsim 30$~GHz for the first epoch, and $\nu_{\rm sa}\sim 7-15$~GHz for the second epoch (see Figures~\ref{fig:patches} and \ref{fig:Fpvsnusa}). In both cases, the low frequency patches are dimmer with respect to the high frequency patches. 

As the peak in the spectrum is not clearly visible during the first epoch, we focus in the following on the second epoch. 
If we assume that the radius of the emitting region is constant, Equation \ref{eq:nusasn}
implies that the magnetic field is a factor of $\sim 1.7$ times larger in the high-frequency patches, than in the low-frequency patches. As the ratio between the values of $F_{\nu_{\rm sa}}/\nu_{\rm sa}^{5/2}$ between the two patches is $\sim 1.4$ (see Figure~\ref{fig:Fpvsnusa}), which implies that $\eta$ increases by a factor of $2.4$ in the high-frequency patches, with respect to the other. In summary, observations are well-fitted by a model in which two regions contribute, with the first one having a magnetic field about $\sim 1.7$ larger, and a surface about $2.4$ times larger. In the case of the first epoch, similar results apply, except the magnetic field ratio is $\gtrsim 3$, and the ratio of the emitting surfaces is $\sim 7$.

The third epoch is well-fitted by  several patches, with nearly constant values of $F_{\nu_{\rm sa}}/\nu_{\rm sa}^{5/2}$ (see the red points in Figure \ref{fig:Fpvsnusa}), extending from $\nu_{\rm sa}=5$ GHz to $\sim $10 GHz. A factor of 2 in frequencies corresponds to a factor of 1.7 in the magnetic field. On the other hand, the fact that the normalized flux is nearly constant implies that 
\begin{eqnarray}
    \eta\propto B
\end{eqnarray}
in this case. That is, regions with larger magnetic field occupy a larger emitting surface.

In the fourth epoch, the number of observations below the self-absorption frequency is insufficient to constrain the structure of the emitting region.

In summary, the magnetic field structure in the emitting region seems to evolve from the first two epochs to the third one. This is consistent with the possible presence of a non-uniform, clumpy density medium at a radius of $\sim 1.10 \times 10^{16}$ cm, as deduced by \citet{nayana2020} from an analysis of the time evolution of the spectra. 
Our results add an additional ingredient to this model, and allow us to constrain the characteristics of the clumps within the stellar wind.

\begin{figure}
\centering
\includegraphics[width=.45\textwidth]{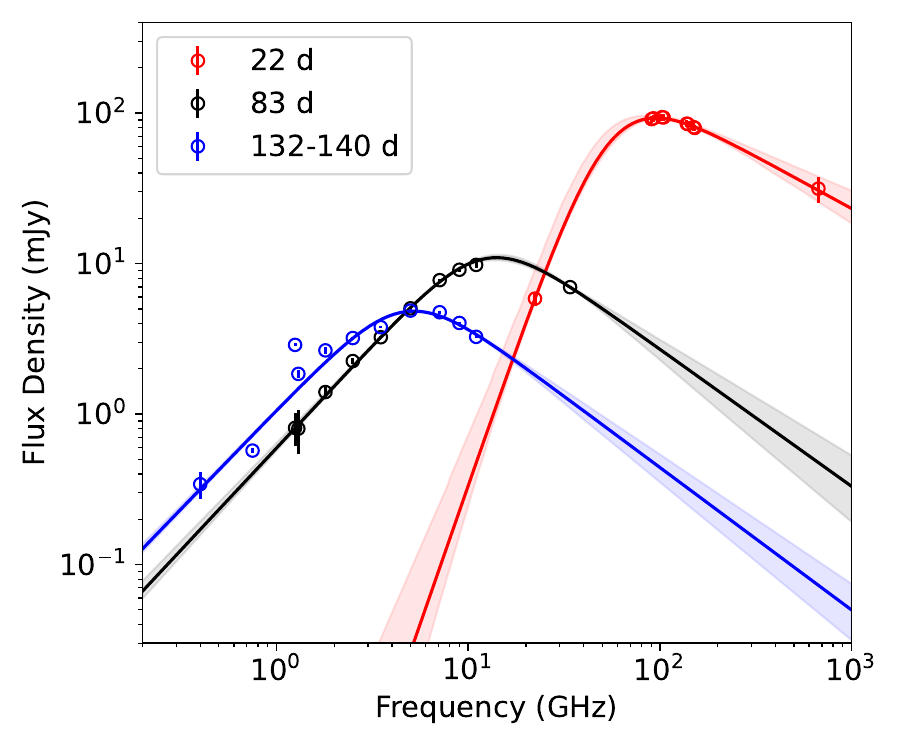}\\
 \caption{Spectrum of the AT2018cow at 22~days, 83 days and 132~days. Data from \citet{Ho2019, Margutti2019, Nayana2021}. Full lines show the fit obtained by using equation \ref{eq:fit}.} 
\label{fig:cowmodel}
\end{figure}

\subsection{Fast blue optical transients}

Fast Blue Optical Transients (FBOTs) are a class of bright cosmic explosions characterised by fast rises and declines in brightness over time scales of a few days. 
Among FBOTs, several events have been studied in detail, with observations extending from radio to X-ray bands: AT2018cow \citep{Ho2019, Margutti2019, Perley2019}, ZTF18abvkwla \citep{Ho2020}, CSS161010 \citep{Coppejans2020}, AT2020xnd \citep{Perley2021, Bright2022} and more recently AT2024wpp \citep{Nayana2025}. 

AT2018cow was discovered on 16 June 2018 in the galaxy CGCG 137-068, at a distance of 60~Mpc \citep{Perley2019}. Radio observations of AT2018cow were obtained with the Karl G. Jansky Very Large Array, very long-baseline interferometry \citep{Margutti2019}, Atacama Large Millimeter/submillimeter Array and the Australia Telescope Compact Array \citep{Ho2019}, and the upgraded Giant Metrewave Radio Telescope 
\citep{Nayana2021}.

Figure \ref{fig:cowmodel} shows the evolution of the radio emission  during three epochs (22, 83 and 132~days), at frequencies ranging from 0.4 to 671~GHz, and a fit to the observations using Equation \ref{eq:fit}.
At $\lesssim$ 22~days, the emission is poorly constrained below the self-absorption frequency, with a power-law slope of $\alpha_1= 3.68 \pm 0.80$ below the self-absorption frequency, which is consistent within $\sim 1.5 \sigma$ with the standard value of 2.5, given the large uncertainties.

At $\gtrsim$ 83~days, the spectrum and light curve evolve dramatically. The peak flux drops from $F_0 = 176.05 \pm 13.34$ mJy at early times to $F_0 = 18.24 \pm 1.59$ mJy at 83~days, and further to $F_0 = 8.87\pm 0.39$ mJy at 132 days. At the same time, the spectrum changes both above and below the observed peak frequency. At low frequencies, the slopes become shallower compared to early times, with $\alpha_1= 1.36\pm 0.05$ at 83~days and $\alpha_1= 1.32\pm 0.03$ at 132 days. Interestingly, the spectral slope above the self-absorption frequency steepens with time (although the errors are relatively large), going from $\alpha_2= -0.70 \pm 0.15$ at 22 days to $\alpha_2= -0.91\pm 0.16$ at 83 days, and $\alpha_2= -0.95\pm 0.09$ at 132 days. 
These slopes correspond to  $p=-2\alpha_2+1=2.40\pm 0.33$, $p=2.82\pm 0.32$, and $p=2.90\pm 0.18$ for the first, second and third epoch respectively. These results indicate that the emission is significantly more complex than a simple SSA spectrum, for which the expected spectral index is $\alpha_\nu=5/2$. 

Given the limited number of data points (in particular, the single frequency at 22~days below the self-absorption frequency), solving the full inverse problem described in Section \ref{model:general} would lead to degenerate results. Therefore, we use a forward modelling approach to gain insights into the inhomogeneity of the system. Furthermore, given the evolution of the spectral slopes and peak fluxes with time, as well as the expected high level of inhomogeneities present in the data, it is reasonable to interpret these inhomogeneities in terms of asymmetries.

In addition, we note that \citet{chevalier98} equations, used to derive the relations presented in this paper, are strictly valid for $\nu_m < \nu_{\rm sa} < \nu_c$, where $\nu_c$ is the cooling frequency, and $\nu_m$ is the characteristic frequency corresponding to electrons with minimum Lorentz factor.
At $t \sim 22$ days, \citet{Ho2019} argued that $\nu_m < \nu_c < \nu_{\rm sa}$, based on the inferred magnetic field and on the relation $\nu_c \propto 1/(B^3 t^2)$. However, this relation assumes that electron cooling time is comparable to the dynamical time, while $\nu_c$ can be larger for electrons that have recently crossed the shock.
\citet{Ho2019} further argued that the \citet{chevalier98} formalism may still be used if shocked electrons are continuously reenergized. In the following, we assume this is the case, and take $\nu_{\rm sa} < \nu_c$. If instead $\nu_c < \nu_{\rm sa}$, the physical parameters deduced at 22 days should be taken as qualitative estimates. A self-consistent study, taking in account the different characteristic frequencies is beyond the scope of this work and is left for a future study.

We start by considering the case where $R=$ constant during each epoch, that is, all the patches are located approximately at the same radial distance $R$. Considering the ratio of the self-absorption frequencies and corresponding fluxes (Equations~\ref{eq:nusa} and \ref{eq:Fsa}) at different epochs, and under the ansatz that the radius scales with time (i.e., nearly constant velocity), we can estimate the magnetic field and covering factor of the patch with the largest $\nu_{\rm sa}$, for the second and third epoch, with respect to the first one. We obtain:
\begin{eqnarray}
    \frac{B(t=22 \;  {\rm days})}{B(t=83 \;  {\rm days})} = 4.3 \qquad
    \frac{\eta(t=22 \;  {\rm days})}{\eta(t=83 \;  {\rm days})} = 87.8\;, \\
    \frac{B(t=22 \;  {\rm days})}{B(t=132 \;  {\rm days})} = 2.7 \qquad
    \frac{\eta(t=22 \;  {\rm days})}{\eta(t=132 \;  {\rm days})} = 997\;.
\end{eqnarray}
Furthermore, we can write the relation between the covering factor and the magnetic field as
\begin{eqnarray}
    \eta \propto B^{\frac{p+6}{p+4}\left(\alpha_1-1-\frac{2p+13}{p+6}\right)} =
    \begin{cases}
\displaystyle B^{-2.27} & {\rm t = 83\; days} \\[8pt]
\displaystyle B^{-2.31}, & {\rm t = 132\; days}
\end{cases}
\label{eq:coweta}
\end{eqnarray}

From these results, we can deduce the following behaviour for the emitting region. During the first epoch, the emission comes from a homogeneous region in terms of magnetic field\footnote{Under the premise that the emission comes from a homogeneous region does not imply that the emitting region is spherical, as it may originate, for instance, from a uniform collimated jet.}. During the second and third epochs, the magnetic field drops by a factor of a few, which, given the relation $B\propto e^{1/2} \propto \rho^{1/2} v$, implies that either or the density of the ambient medium is smaller by a factor of $\sim 10$, or the velocity is dropping substantially (by a factor of a few), or a combination of both. On the other hand, the size of the patch with this magnetic field intensity decreases dramatically, by a factor of $\sim 100$ from the first to the second epoch, and by $\sim 1000$ from the first to the third epoch. Consistent with Equation~\ref{eq:coweta}, during the second and third epochs most of the emitting surface has low or intermediate magnetic fields, while large magnetic fields are limited to a small fraction of the emitting region. 

In summary, while in the first epoch the magnetic field is large and homogeneous, it becomes patchy and dominated by low intensities at later times. 
This behaviour is consistent with an outflow that is switched off and spreads laterally, 
and it is supported by extensive observations at optical, X-ray, millimeter, and radio wavelengths that have constrained the progenitors of these events. The central engine has been identified as a compact object \citep{Margutti2019}, producing a mildly relativistic shock (with velocities v$\sim 0.1$–$0.2$ c) propagating through a dense circumstellar material \citep{Ho2019}. 
Multi-wavelength follow-up also showed evidence for strong asymmetries in the ejecta of AT2018cow \citep{Margutti2019}, based on different velocity regimes of velocity of the ejecta from $v=0.1-0.3c$. This scenario is also supported by polarisation measurements \citep{maund2023}
between $5$ and $8$~days after the explosion, which suggests an asymmetric flow and likely the presence of a jet.

The case of $R$ not being constant is more complex, as it introduces an additional degree of freedom. 
We can generalize Equation~\ref{eq:coweta} by assuming $R\propto B^b$. Then, we obtain
\begin{eqnarray}
    \eta \propto B^{\frac{2b+p+6}{p+4}\left(\alpha_1-1-\frac{2p+13}{p+6}\right)} =
    \begin{cases}
\displaystyle B^{-0.52 b - 2.26} & {\rm t = 83\; days} \\[8pt]
\displaystyle B^{-0.68 b - 2.31}, & {\rm t = 132\; days}
\end{cases}\end{eqnarray}
For $b=0$, we recover Equation~\ref{eq:coweta}. Cases with $b > -4.4$ correspond, during both epochs, to emitting regions where the magnetic field is larger at larger radii (for example, an outflow structured along the radial direction). In these cases, $\eta$ becomes steeper than in the $b=0$ scenario if $b>0$, and less steep if $-4.4 < b < 0$. For $b< -4.4$, $\eta$ increases with $B$. This case would correspond to patches with large magnetic fields localized at small radii (for instance, an outflow interacting with dense material in the equatorial plane while expanding into low-density material along the polar direction).

\section{Discussion}
\label{sec:discussion}

Radio spectra of some SNe, as well as essentially all FBOTs with well sampled radio spectra, among other high energy transients, frequently show, around the self-absorption frequency, a spectrum broader than expected \citep[e.g.][]{chandra2024b, soderberg2005, bjornsson2013, bjornsson2017, chandra2019, nayana2020, sfaradi2024, nayana2025b}, as well as deviations from the $\nu^{5/2}$ predicted at frequencies below the self-absorption frequency 
(e.g., \citealt{Margutti2019, nayana2020}, Wiston et al., in preparation) and deviations from the $\nu^{-(p-1)/2}$ slope predicted at frequencies above the self-absorption frequency \citep[e.g.,][]{chandra2024b}. 
These features have been interpreted \citep{bjornsson2013, bjornsson2017} as evidence of inhomogeneities in the emitting region.

Previous works \citep[see e.g.][]{bjornsson2013, bjornsson2017} have modelled the presence of inhomogeneities in the emitting region by assuming the magnetic field changes. These authors described the magnetic field distribution in terms of a covering factor, which represents the fraction of the emitting region occupied by gas with a certain value of the magnetic field, and assume a power-law distribution for it. 

Nevertheless, such an approach has two limitations: the first one is that the emitting region depends on other physical parameters, in addition to the magnetic field (e.g., the radius and the microphysical parameters of the accelerated electrons). The second one is that assuming a power-law dependence for the magnetic field distribution does not necessarily represent the true structure of the emission region. 

In this paper, we extended previous methods, by interpreting the inhomogeneities in terms of patches, with each patch described completely by its self-absorption frequency and peak frequency (once the value of $p$ is fixed from observations). We have shown that the key parameter determining the structure of the emission region is $F(\nu_{\rm sa})$, representing the peak flux corresponding to patches with a certain value of the self-absorption frequency. The problem can be represented as an inverse problem, from which, using standard inversion methods \citep[see e.g.,][]{decolle08, decolle10}, $F(\nu_{\rm sa})$ can be determined if a good frequency sampling is available below the self-absorption frequency, as the maximum number of patches that can be recovered is limited by the number of observations.
 
Basically, we have divided the process into two parts: in the first one, the function $F(\nu_{\rm sa})$ can be found by following a rigorous mathematical approach, which depends only on observational quantities. In the second one, physical parameters can be inferred only once this distribution has been determined precisely. As the synchrotron self-absorption frequency depends on several physical parameters, including the magnetic field strength, the size of the emitting region, and the microphysical parameters, inferring physical parameters from $F(\nu_{\rm sa})$ is a degenerate process. Nevertheless, in some cases it is reasonable to assume that the emitting region comes from approximately the same radius, and that the microphysical parameters are the same for the different patches. Then, it is possible to infer the sizes of the emitting region corresponding to each magnetic field intensity, and to get a detailed description of the physical conditions of the emitting region. 

In the case of a power-law dependence for  $F(\nu_{\rm sa}) = k \nu_{\rm sa}^\alpha$, we have shown that the observed flux follows $F_\nu \propto \nu^{\alpha+1}$ for $\alpha<3/2$ and $F_\nu \propto \nu^{5/2}$ for $\alpha>3/2$. Then, values larger than $3/2$ cannot be recovered from the low-frequency spectral slope, but still can produce a broadening of the spectrum around $\nu_{\rm sa}$. This implies that a correlation between $F(\nu_{\rm sa})$ and $\nu_{\rm sa}$ is needed, to get changes in the low-frequency spectral slope, while uncorrelated dependencies lead only to a broadening of the spectrum 

While we have shown that inverse methods can help to gain insights on the structure of the emitting region from unresolved observations, this method has strong limitations, as it cannot distinguish the exact spatial distribution of each patch, and it cannot distinguish between inhomogeneity and asymmetry. That is, when a certain distribution of patches has been recovered, it is not possible to determine whether such a distribution has for instance an angular dependence (which implies that the source is asymmetric in the angular direction), or if the distribution of patches is chaotic (then, associated for instance to turbulence in the post-shock region). Nevertheless, it can be argued that evolution of the degree of inhomogeneity as a function of time strongly suggests the presence of large-scale asymmetries.

Different possibilities have been suggested \citep{bjornsson2013, bjornsson2017, bjornsson2024} to explain inhomogeneities in the emitting regions. These are basically related to inhomogeneities in the particle acceleration process, or inhomogeneities or asymmetries in the emitting region, including instabilities leading to variations in the magnetic field, presence of clumps in the wind of the progenitor star, or asymmetries in the ejecta or environment.

Finally, we mention other possible effects that can affect the low-frequency spectrum. The first one is interstellar scintillation, which can be important at low-frequencies, leading to a systematic error in the flux,  especially when only a few data points are available. The second one is free-free absorption, which can be important in SNe type IIn, in which the density of the ambient medium is large enough. 

\section{Conclusions}
\label{sec:summarize}

In this paper, we presented a method to infer information about the inhomogeneities and asymmetry of astrophysical sources from unresolved self-absorbed radio spectra. We showed that, in general, the degree of inhomogeneity in the emitting region determines the slope of the low-frequency spectrum and the flux around the self-absorption frequencies. Inhomogeneous media result naturally in low-frequency slopes shallower than the standard $\nu^{5/2}$ synchrotron spectrum and/or in a broadening of the spectrum at frequencies close to the self-absorption frequencies.

We applied the method to SN 2016coi, treating directly the general inverse problem; and to FBOT AT2018cow, by using a more standard forward approach. In both cases, we showed that the emitting region is inhomogeneous. While in the case of SN 2016coi the inhomogeneity is limited, for FBOT AT2018cow it is dramatic, implying a very asymmetric and evolving emitting region. 

We have shown clearly the degrees of degeneracy intrinsically present in the data, which could and should be taken into account when building forward models of radio sources. Although these degeneracies imply that the inversion process is typically non-unique, the level of degeneracy present in the system can be inferred from the methods presented here. 

The methods developed in this paper are applicable to radio observations of supernovae, FBOT, tidal disruption events, and gamma-ray bursts, although the relativistic motion of these sources make the method more complex. In GRBs, asymmetries are often treated by considering structured jets during the prompt of afterglow phases \citep[see, e.g,][and references therein]{salafia2022}. Inverse methods have been applied to the reconstruction of the structure of the GRB 170817A from its afterglow emission \citep{takahashi2020, takahashi2021}.

The results presented in this paper outline the importance of getting very well sampled, multi-epoch spectra (currently available only in a handful of cases for, e.g., SNe). 
The order of magnitude sensitivity improvement promised by the next generation of radio telescopes, e.g., the SKA (operating from 50\,MHz to 14\,GHz) and the ngVLA (operating from ${\sim}1$ to $100$\,GHz), will allow the observing bandwidths to be split with finer frequency resolution, which in turn will provide radio spectra with tens of data points below the self-absorption frequencies. 
These future broadband observations finely sampled across a large range of frequencies will allow us to infer much better information on the structure of the emitting regions. 

\section*{Acknowledgements}

We acknowledge the computing time granted by DGTIC UNAM on the supercomputer Miztli (project LANCAD-UNAM-DGTIC-281). 
We acknowledge support from the DGAPA/PAPIIT grant IN113424. GU acknowledges support from University of California-Alianza MX Research \& Innovation: UCPRF2025-01. GU also acknowledges the support of the Heising-Simons Foundation and the Vera Rubin Presidential Chair at the University of California, Santa Cruz.
JKL acknowledges support from the University of Toronto and Hebrew University of Jerusalem through the University of Toronto--Hebrew University of Jerusalem Research and Training Alliance program. 
The Dunlap Institute is funded through an endowment established by the David Dunlap family and the University of Toronto. 
LGG gratefully acknowledge the support of the Simons Foundation (MP-SCMPS-00001470, N.G.). The authors used AI-based tools for English grammar correction. All scientific content was developed by the authors.

\section*{Data Availability}
The data underlying this article will be shared on reasonable request to the corresponding author.

\bibliographystyle{mnras}
\bibliography{references}


\appendix
\section{Inferring the radius of the emitting region}
\label{ap:A}

In the standard \citet{chevalier98} model, the shock radius is obtained from Equations~\ref{eq:nusa} and \ref{eq:Fsa}, as 
\begin{eqnarray}
  R \propto F_{\nu_{\rm sa}}^{(p+6)/(2p+13)} \nu_{\rm sa}^{-1}\;.
\end{eqnarray}

In the inhomogeneous case, part of the flux is redistributed around the self-absorption frequency, leading to a reduction of the peak flux. Then, the radius inferred when assuming the emitting region is homogeneous underestimates the true value by a factor 
\begin{eqnarray}
  \frac{R_{\rm inh}}{R_{\rm hom}} = \left(\frac{F_{\nu_0,\rm inh}}{F_{\nu_0,\rm hom}}\right)^{(p+6)/(2p+13)}\;.
\end{eqnarray}

The observed flux can be written as 
\begin{eqnarray}
    F_\nu = \int_{\nu_{\rm sa,min}}^{\nu_{\rm sa,max}} F(\nu_{\rm sa}) \left(\frac{\nu}{\nu_{\rm sa}} \right)^{5/2} \left[1-e^{-(\nu/\nu_{\rm sa})^{-(p+4)/2}} \right]
    d\nu_{\rm sa}\;.
\end{eqnarray}

In the homogeneous case, we have 
$F(\nu_{\rm sa}) = k_1 \delta(\nu_{\rm sa}-\nu_0)$
and the emitted flux reduces to 
\begin{eqnarray}
    F_\nu = k_1 \left(\frac{\nu}{\nu_0} \right)^{5/2} \left[1-e^{-(\nu/\nu_0)^{-(p+4)/2}} \right]\;.
\end{eqnarray}
The value of the normalization constant $k_1$ is obtained from
\begin{eqnarray}
  k_1 = \int_{\nu_{\rm sa,min}}^{\nu_{\rm sa,max}} F(\nu_{\rm sa}) d\nu_{\rm sa}
\end{eqnarray}
Then, the flux ratio is 
\begin{eqnarray}
  \frac{F_{\nu_0,\rm inh}}{F_{\nu_0,\rm hom}} = \frac{    \int_{\nu_{\rm sa,min}}^{\nu_{\rm sa,max}} F(\nu_{\rm sa}) \left(\frac{\nu}{\nu_{\rm sa}} \right)^{5/2} \left[1-e^{-(\nu/\nu_{\rm sa})^{-(p+4)/2}} \right]
    d\nu_{\rm sa}
}{\int_{\nu_{\rm sa,min}}^{\nu_{\rm sa,max}} F(\nu_{\rm sa}) d\nu_{\rm sa} \left(\frac{\nu}{\nu_0} \right)^{5/2} \left[1-e^{-(\nu/\nu_0)^{-(p+4)/2}} \right]}
\label{eq:a6}
\end{eqnarray}

This Equation is general, and can be evaluated numerically once $F(\nu_{\rm sa})$ is obtained (e.g., by an inverse method). For simplicity, in the following we consider the case $F(\nu_{\rm sa}) = k_2 \nu_{\rm sa}^\alpha$ for $\nu_{\rm sa,min} < \nu_{\rm sa} < \nu_{\rm sa,max}$,
and $F(\nu_{\rm sa})=0$ otherwise. 
In this case, the normalization factor becomes 
\begin{eqnarray}
  k_1 = k_2 \int_{\nu_{\rm sa,min}}^{\nu_{\rm sa,max}} \nu_{\rm sa}^\alpha d\nu_{\rm sa} = k_2 \frac{\nu_{\rm sa,max}^{\alpha + 1} - \nu_{\rm sa,min}^{\alpha + 1}}{\alpha+1}.
\end{eqnarray}
Considering $\nu_{\rm sa,max}\gg \nu_{\rm sa,min}$, we have two cases: if $\alpha > -1$, we get
$k_1 \simeq k_2 \nu_{\rm sa,max}^{\alpha + 1}/(\alpha+1)$, while if $\alpha < -1$, we get
$k_1 \simeq -k_2 \nu_{\rm sa,min}^{\alpha + 1}/(\alpha+1)$.

Then, the flux ratio becomes 
\begin{eqnarray}
    \frac{F_{\nu_0,\rm inh}}{F_{\nu_0,\rm hom}} = 
\frac{\int_{\nu_{\rm sa,min}}^{\nu_{\rm sa,max}} k_2 \nu_{\rm sa}^\alpha \left(\frac{\nu_0}{\nu_{\rm sa}} \right)^{5/2} \left[1-e^{-(\nu_0/\nu_{\rm sa})^{-(p+4)/2}} \right]
    d\nu_{\rm sa}}{k_1 (1-e^{-1})
}\;.
\end{eqnarray}
Setting $x=\nu/\nu_{\rm sa}$, and defining 
\begin{eqnarray}
    I(a,b) = \int_a^b x^{1/2-\alpha} \left[1-e^{-x_0^{-(p+4)/2}} \right] dx\;,
\end{eqnarray}
we get
\begin{eqnarray}
    \frac{F_{\nu_0,\rm inh}}{F_{\nu_0,\rm hom}} = 
\frac{(\alpha+1) \nu_0^{\alpha+1} I(\nu_0/\nu_{\rm sa,max},\nu_0/\nu_{\rm sa,min})}{ (\nu_{\rm sa,max}^{\alpha + 1} - \nu_{\rm sa,min}^{\alpha + 1})(1-e^{-1})
}\;.
\label{eq:ratio}
\end{eqnarray}
If $\alpha > 0$, then $\nu_0 \simeq \nu_{\rm sa,max}$, while if $\alpha < 0$, $\nu_0 \simeq \nu_{\rm sa,min}$. Then, 
\begin{eqnarray}
\frac{F_{\nu_0,\rm inh}}{F_{\nu_0,\rm hom}} =
\frac{\alpha+1}{1-e^{-1}}
\begin{cases}
\displaystyle I(1,\infty), & \alpha > 0. \\[8pt]
\displaystyle \left(\frac{\nu_0}{\nu_{\rm sa,max}}\right)^{\alpha+1} I(0,1), & -1 < \alpha < 0. \\[8pt]
\displaystyle I(0,1), & \alpha < -1.
\end{cases}
\end{eqnarray}

\bsp	
\label{lastpage}
\end{document}